\vskip5pt
\centerline {\bf Discrete symmetries and quantum number conservation}
\vskip10pt
\centerline {Douglas Newman}
\vskip5pt
\centerline {e-mail: \it dougnewman276@gmail.com}
\vskip 10pt

\beginsection Abstract 

The algebraic formulation of discrete $P$ and $T$ space-time 
symmetries is related to fermion quantum numbers defined by a 
$Cl_{3,3}$ sub-algebra of the $Cl_{7,7}$ Clifford Unification 
algebra. Fermion decays and interactions have been shown to 
conserve all seven binary quantum numbers defined by $Cl_{7,7}$.
The previously formulated {\it Conservation Law} is modified
to include the effects of employing distinct F,G quantum 
numbers in descriptions of fermions with C=+1 and C=$-1$.
This is relevant in interpreting the results of high energy 
experiments. 

\vskip20pt 

\beginsection \S1 Introduction 

Clifford Unification (CU) [1,2] provides a seven binary quantum number 
description of fermions, related to their physical properties.  
Each quantum number is defined by one of seven commuting elements of 
$Cl_{7,7}$, providing unique physical interpretations of all the
elements of this algebra. One of the seven quantum numbers
was identified in [2] as fermion intrinsic parity, leading to the expectation that
this is preserved in all fermion interactions. This conflicts
with experimental evidence for the non-conservation of parity that is
now generally accepted and reported in all the textbooks on particle
physics (e.g.[3]). This apparent conflict in the interpretation of experimental
results appears to be a consequence of the different algebraic definitions of parity 
in CU[1,2] and the Standard Model that are described in [4].

Sections 2 and 3 are concerned with establishing the algebraic relationship
between the $Cl_{3,3}$ and classical descriptions of parity and time-reversal
symmetries. \S2 gives the coordinate descriptions
of parity inversion ($P$) and time-reversal ($T$) symmetries in terms of the 
space-time algebra $Cl_{1,3}$. \S3 identifies $P$ and $T$ symmetries with
quantum numbers determined by commuting elements of $Cl_{3,3}$ . 
\S4 defines the ABCDEFG quantum number description of fermions. It highlights 
the parity (C) dependence of F and G, which distinguish fermion generations, 
correcting the results in \S8 of [2] and earlier versions of this paper. 
\S5 formulates a modified conservation law of all seven quantum
numbers in elementary particle decays and interactions. This relates to the
table of fermion quantum numbers given in the Appendix. 

\beginsection \S2. Algebraic expressions for discrete space-time symmetries in classical physics

Parity inversion and time-reversal ({\it P} and {\it T}) are discrete transformations, 
conventionally defined in terms of sign changes in the space-time 
coordinates $ x_\mu \{\mu = 0,1,2,3\}$, viz.
$$
P: x_\mu \to -x_\mu\> {\rm for}\> \{\mu=1,2,3\};\>\> T: x_0 \to -x_0. \eqno(2.1)
$$
Changing the sign of a single spatial coordinate also changes the parity of the
coordinate system as a rotation of the other two coordinates through $180^0$ 
changes the sign of both. All rotations of spatial coordinate
systems can be expressed as Lorentz transformations, which leave their parity unchanged.

4-vectors describing space-time displacements $\Delta x_\mu$ of a particle
(expressed in terms of an arbitrary Minkowski reference frame)  
have the algebraic form
$$
\Delta {\bf s} = \gamma^s\Delta s = \gamma^\mu\Delta x_\mu.            \eqno(2.2)
$$ 
Here the $\gamma^\mu$ are anti-commuting components of a covariant 
Minkowski tensor, related to the contravariant components
$\gamma^0 = \gamma_0,\>{\rm and} \>\gamma^\mu =-\gamma_\mu\> {\rm if}\> \{\mu =1,2,3\}$.
Together, the $\gamma_\mu$ generate a $Cl_{1,3}$ Clifford algebra, isomorphic 
to the Dirac matrix algebra, but with a different physical interpretation. 
$\gamma^0$ is interpreted as a unit displacement in time,
and the $\gamma^\mu \{\mu=1,2,3\}$ are interpreted 
as unit displacements in the three orthogonal directions
in space. $\Delta {\bf s} $ is the change in `proper' time, as 
measured on a (conceptual) clock at rest with respect to
the particle and $\gamma^s$ (where $(\gamma^s)^2 = 1 $) is 
the unit proper time. The Lorentz transformation 
ensures that $\Delta s$ and $\Delta x_0$ have the same sign.

{\it P} and {\it T} can be expressed algebraically as 
similarity transformations of all the $\gamma_\mu$, viz.
$$\eqalign{
P: \gamma^\mu \to \gamma^0 \gamma^\mu (\gamma^0)^{-1}=&\>\gamma^\mu = -\gamma_\mu \>\>{\rm for}\> \{\mu =1,2,3\}
\>{\rm or}\>+\gamma^\mu \>{\rm for}\> \{\mu = 0\},\cr
T: \gamma^\mu \to {\bf v} \gamma^\mu (\bf v)^{-1}=& +\gamma^\mu \>\>{\rm for}\> \{\mu =1,2,3\}
\>{\rm or}\>-\gamma^\mu \>{\rm for}\> \{\mu = 0\}.}                      \eqno(2.3)
$$
where ${\bf v}=\gamma^1\gamma^2\gamma^3$ is the algebraic expression for unit spatial volumes. 
Defining $\gamma^\pi= \gamma^0{\bf v} =\gamma^0\gamma^1\gamma^2\gamma^3$, 
the combined $PT$ transformation is
$$
PT: \gamma^\mu \to \gamma^\pi \gamma^\mu (\gamma^\pi)^{-1}= -\gamma^\mu\> {\rm all}\> \mu,\>\> {\rm and}\> 
\gamma^s \to \gamma^\pi \gamma^s (\gamma^\pi)^{-1} = -\gamma^s,           \eqno(2.4)
$$ 
showing that $PT$ produces {\it proper time} reversal. 
As the above analysis is necessarily confined to space-time geometry, it
does not provide a description of charge conjugation.
It is of interest, nevertheless, that electric current 4-vectors ${\bf j}= j^\mu \gamma_\mu$
are invariant under the $PT$ transformation {\it and} a change in sign of the charge.
\vskip10pt

\beginsection \S3.  $Cl_{3,3}$ description of space-time symmetries in quantum mechanics

In CU discrete symmetries are identified with the eigenvalues
of specific commuting elements of the algebra. This provides unique 
identifications, which are preferable to descriptions based 
on transformations of wave-functions. The physical interpretation of 
the $Cl_{3,3}$ algebra has been given in [1,2]. It has six generators:
$\hat\gamma^1,\>\hat\gamma^2,\>\hat\gamma^3$ which correspond to the
unit elements in the three spatial directions, and 
$\hat\gamma^a, \hat\gamma^b,\>\hat\gamma^c$ which are Lorentz
invariants describing fermions and their interactions. 
Unit elements of time are defined as the product
$\hat\gamma^0 =\hat\gamma^1\hat\gamma^2\hat\gamma^3\hat\gamma^a\hat\gamma^b\hat\gamma^c$.
Equation (2.2) for a fermion displacement becomes
$$
\Delta {\bf s} = \hat\gamma_{*0}\Delta x^{*0} = \hat\gamma_\mu\Delta x^\mu.            \eqno(3.1)
$$
where $\hat\gamma_{*0}$ is the unit of proper time measured in the fermion rest frame.
More generally starred coordinates indicate that unit space-time displacements referred 
to the fermion rest frames. Ordered products of generators are abbreviated, for example 
$\hat\gamma^{*2}\hat\gamma^6 =\hat\gamma^{*2a} $.

$Cl_{3,3}$ has, at most, seven commuting elements, for example
$\hat\gamma^{*13},\>\hat\gamma^{*2a},\>\hat\gamma^{bc}, $
$\>\hat\gamma^{*0}, \hat\gamma^{*0bc},\>\hat\gamma^{*031},\>\hat\gamma^{*02a}$.
Any three of these that are not related by multiplication
take eigenvalues that distinguish the eight leptons, the simplest physical interpretation being
obtained with $\hat\gamma^{*13},\>\hat\gamma^{*0},\>\hat\gamma^{bc}$. $\hat\gamma^{*2}$ 
is the spatial orientation of the fermion spin $\hat\gamma^{*13}$. 
Changes in parity are produced by the coordinate transformations
$$\eqalign{
	\hat\gamma^{*2b}\hat\gamma^{*1}\hat\gamma^{*2b}=&\hat\gamma^{*1},\>\>\>
	\hat\gamma^{*2b}\hat\gamma^{*2}\hat\gamma^{*2b}=-\hat\gamma^{*2},\>\>\>
	\hat\gamma^{*2b}\hat\gamma^{*3}\hat\gamma^{*2b}=\hat\gamma^{*3},\>\>\>
	\hat\gamma^{*2b}{\bf v}\hat\gamma^{*2b}=-{\bf v}, \cr
	\hat\gamma^{*2c}\hat\gamma^{*1}\hat\gamma^{*2c}=&\hat\gamma^{*1},\>\>\>
	\hat\gamma^{*2c}\hat\gamma^{*2}\hat\gamma^{*2c}=-\hat\gamma^{*2},\>\>\>
	\hat\gamma^{*2c}\hat\gamma^{*3}\hat\gamma^{*2c}=\hat\gamma^{*3},\>\>\> 
	\hat\gamma^{*2c}{\bf v}\hat\gamma^{*2c}=-{\bf v}, \cr }           \eqno (3.2)
$$
both of which satisfy $(\hat\gamma^{*2b})^2 = (\hat\gamma^{*2c})^2  = 1$. It follows that
(3.2) describes similarity transformations changing the sign of $\hat\gamma^{*2} $. 
Similarly
$$
\hat\gamma^{*2b}\hat\gamma^{bc}\hat\gamma^{*2b}=-\hat\gamma^{bc},\>\>\>
\hat\gamma^{*2c}\hat\gamma^{bc}\hat\gamma^{*2c}=-\hat\gamma^{bc},            \eqno (3.3)
$$
changing the sign of $\hat\gamma^{bc}$.
Equations (3.2) and (3.3) still hold if the coordinate direction 
 $\hat\gamma^{*2}$ is changed throughout into $\hat\gamma^{*1}$
or $\hat\gamma^{*3}$, showing that {\it any} change in parity produces
a change in the sign of $\hat\gamma^{78}$.
Furthermore, as equations (3.2) and (3.3) depend only on the multiplication 
properties of the $\hat\gamma$s, they hold in any Lorentz frame, always producing 
a change in sign of a single spatial coordinate direction, and a consequent change
in the parity of the coordinate system. $\hat\gamma^{bc}$
is Lorentz invariant, changing sign for any choice of coordinates, 
and its eigenvalues therefore determine both coordinate parity 
and the intrinsic parity of leptons.

Given that the eigenvalues of $\hat{\gamma}^{*0}$ determine the time direction
in the rest frame of the fermion, time reversal must be expressed in terms of
an element of the algebra that anti-commutes with $\hat{\gamma}^{*0}$ and
commutes with the unit spatial displacements 
$\hat\gamma^{*1},\>\hat\gamma^{*2},\>\hat\gamma^{*3}$. There are
two possibilities, viz. $\hat\gamma^{*06}$ and
$\hat\gamma^{*\pi 0} = \hat\gamma^{*1}\hat\gamma^{*2}\hat\gamma^{*3}$.
These satisfy $(\hat\gamma^{*0a})^2 = -1$ and $(\hat\gamma^{*\pi 0})^2 = 1$.
The most satisfactory choice is $\hat\gamma^{*\pi 0}$, which does not involve
time-like generators and can be expressed as the similarity transformation
$$
\hat\gamma^{*0}\to (\hat\gamma^{*\pi 0})^{-1}\,\hat\gamma^{*0}\,\hat\gamma^{*\pi 0} 
= -\hat\gamma^{*0}.                        \eqno (3.4) 
$$
The identification of these symmetries with quantum numbers ensures their invariance
in physical processes. Nevertheless, while (3.4) holds in any reference frame, only
the time direction $\hat\gamma^{*0}$ in the fermion rest frame defines a Lorentz
invariant quantum number.

\beginsection \S4. Quantum number description of fermions 

It was shown in [1,2] that each of the $2^7$ fermion states is specified
by a unique combination of the seven binary quantum numbers A, B, C, D, E, F, G, 
(denoted $\mu_A$, etc. in [1]), which will be referred to as their `signatures'. 
A=$\pm 1$ specifies spin direction, the eigenvalues B of $\hat\gamma^{*0}$
distinguish fermions (B=1) from anti-fermions (B=$-1$), so that A, B determine the
four {\it states} of any fermion. The eigenvalues (C=$\pm1$) 
of $\hat\gamma^{78}$ determine their intrinsic parity, distinguishing the two 
states in all fermion doublets. D and E, together, determine the three colours
of quarks, and separate quarks and leptons.  The F$_{\rm _C}$,G${\rm _C}$
quantum numbers, which distinguish fermion generations, also depend on C=$\pm1$.
This is made explicit in the table of fermion signatures given in the Appendix.
F$_{\rm _C}$,G${\rm _C}$ descriptions of the 
the u,c,t (C=$-1$) quarks and d,s,b (C=+1) quarks are related by
the CKM matrix (as in the Standard Model). Similarly, descriptions
of the e$^-$, $\mu^-$, $\tau^-$ (C=1) leptons and $\nu_e$, $\nu_\mu$, $\nu_\tau$
(C=$-$1) neutrinos are related by the PMNS matrix (as in the Standard Model).

The CKM matrix (e.g. see [3] \S14.1) is unitary and has numerical components, viz.
$$
{\bf V}_{\rm CKM} = \left(\matrix{\rm V_{ud} &\rm V_{us}&\rm V_{ub}\cr 
	\rm V_{cd} &\rm V_{cs}&\rm  V_{cb}\cr
    \rm V_{td} &\rm V_{ts}&\rm V_{tb}\cr}\right), \eqno (4.1)
$$
or its inverse
$${\rm
	{\bf V}^*_{\rm CKM} = \left(\matrix{\rm V^*_{ud} &\rm V^*_{cd}& \rm V^*_{td}\cr 
		\rm V^*_{us} &\rm V^*_{cs}& \rm V^*_{ts}\cr
		\rm V^*_{ub} &\rm V^*_{ts}& \rm V^*_{tb}\cr}\right)},\eqno (4.2)
$$
where the starred components are complex conjugates. These are interpreted in
this work as relating the (F$_-$G$_-$) quantum numbers of the C=$-1$ u,c,t quarks
and the (F$_+$ G$_+$) quantum numbers of the C=+1 d,s,b quarks.
$$\eqalign{
	{\rm u}(1_- 1_-)&=\rm V_{ud}d(1_+ 1_+ ) +V_{us}s(1_+ \bar1_+ )+ V_{ub} b(\bar1_+ 1_+ )\cr
	{\rm c}(1_- \bar1_-)&=\rm V_{cd}d(1_+ 1_+ ) + V_{cs}s(1_+ \bar1_+ ) + V_{cb} b(\bar1_+ 1_+ )\cr
	{\rm t}(\bar1_- 1_-)&=\rm V_{td}d(1_+ 1_+ ) +V_{ts}s(1_+ \bar1_+ ) + V_{tb} b(\bar1_+ 1_+ )\cr}.\eqno (4.3)
$$ 
The corresponding inverse relations are 
$$\eqalign{
	{\rm d}(1_+ 1_+ )&=\rm V^*_{ud}u(1_- 1_-) +V^*_{cd} c(1_- \bar1_-) + V^*_{td}t(\bar1_- 1_-)\cr
	{\rm s}(1_+ \bar1_+ )&=\rm V^*_{us} u(1_- 1_-)  +V^*_{cs} c(1_- \bar1_-) + V^*_{ts}t(\bar1_- 1_-)\cr
	{\rm b}(\bar1_+ 1_+ )&=\rm V^*_{ub} u(1_- 1_-) +V^*_{cb}c(1_- \bar1_-) +V^*_{tb}t (\bar1_- 1_-)\cr} \eqno (4.4)
$$ 
Magnitudes of the elements of the CKM matrix have been determined experimentally 
(e.g \S14.3 of [3]):

$$
{\bf |V|}_{\rm CKM} = \left(\matrix{\rm |V_{ud}| &\rm |V_{us}|&\rm |V_{ub}|\cr 
	\rm |V_{cd}| &\rm |V_{cs}|&\rm  |V_{cb}|\cr
	\rm |V_{td}| &\rm |V_{ts}|&\rm |V_{tb}|\cr}\right) = 
\left(\matrix{0.974 &0.225&0.004\cr 0.225&0.973&0.041\cr 0.009 &0.040&0.999\cr}\right)  \eqno (4.5)
$$
In the following discussion it will be assumed that ${\bf V}_{\rm CKM}$
has real components.

\vskip 10pt

\beginsection \S5. Quantum number conservation 

Descriptions of {\it corresponding} anti-fermions 
are obtained by changing the sign of all seven fermion quantum numbers 
(labelled  A B C D E F G). Although fermions can change in interactions, 
quantum numbers do not change, giving the 
\vskip 5pt
 {\bf Conservation Law:} {\it All seven quantum numbers are conserved in fermion decays and interactions}
\vskip5pt
It is convenient to introduce a condensed notation in demonstrating 
applications of this law, writing $\bar1$ for $-1$ in fermion descriptions.
Following the table given in
the Appendix, F$_\pm$, G$_\pm$ quantum numbers for C=1 and C=$-1$
fermions are distinguished by attaching + or $-$
suffices to give $1_+$, $1_-$ and $\bar1_+$, $\bar1_-$ respectively.
For example, electrons of either spin are described 
by the set of six quantum numbers  B C D E F G as 
$\{{\rm e}^-:1\>1\>\bar1\>\bar1\>1_+\>1_+\>\}$. 
Similarly, neutrinos of either spin are described
by $\{{\nu_e}:1\>\bar1\>\bar1\>\bar1\>1_-\>1_-\>\}$

Simple examples are provided by decay processes.
$\beta$ decay is produced by the first generation process 
$ {\rm d}\to {\rm u}+ {\rm e}^- + \nu_e$, 
corresponding to the equation
$$ 
\{{\rm d}_b:1\>1\>1\>1\>1_+\>1_+\}=\{{\rm u}_b:1\>\bar1\>1\>1\>1_-\>1_-\} 
+\{{\rm e}^-:1\>1\>\bar1\>\bar1\>1_+\>1_+\}+\{\bar\nu_e:\bar1 \>1\>1\>1\>\bar1_-\>\bar1_-\},    \eqno (5.1)
$$
where the equality can be checked by adding each of the six corresponding numbers
in the brackets on the right hand side of the equation. Different quark colours change
the signs of D and E in the same way for both d and u, leaving the equality the same.
The F,G quantum numbers cancel for both C=1 and C=$-1$ leptons, so complications produced 
by the CKM matrix do not occur, and the simple form of conservation law 
(above )is maintained.
 
Analogous equations also hold for second and third generation $\beta$ decays, viz.
$$\eqalign{ 
\{{\rm s}_b:1\>1\>1\>1\>1_+\>\bar1_+\}=&\{{\rm c}_b:1\>\bar1\>1\>1\>1_-\>\bar1_-\} 
+\{\mu^-:1\>1\>\bar1\>\bar1\>1_+\>\bar1_+\}+\{\bar\nu_\mu:\bar1 \>1\>1\>1\>\bar1_-\>1_-\},\cr 
\{{\rm b}_b:1\>1\>1\>1\>\bar1_+\>1_+\}=&\{{\rm t}_b:1\>\bar1\>1\>1\>\bar1_-\>1_-\} 
+\{\tau^-:1\>1\>\bar1\>\bar1\>\bar1_+\>1_+\}+\{\bar\nu_\tau:\bar1 \>1\>1\>1\>1_-\>\bar1_-\}.}   \eqno (5.2)
$$
 
Corresponding scattering processes is described by moving one set of fermion quantum numbers 
to the other side equations of equations (5.1) and (5.2). For example, (5.1) becomes 
$$ 
\{{\rm d}_b:1\>1\>1\>1\>1_+\>1_+\}+\{\nu_e:1\>\bar1\>\bar1\>\bar1\>1_-\>1_-\}
=\{{\rm u}_b:1\>\bar1\>1\>1\>1_-\>1_-\}+\{{\rm e}^-:1\>1\>\bar1\>\bar1\>1_+\>1_+\},             \eqno (5.3)
$$
which describes the scattering of the $\nu_e$ neutrino by a d quark to produce
a u quark and an electron. 

The same relation also describes the $\beta$ decay of quarks into leptons, e.g.
$$
\{{\rm d}_b:1\>1\>1\>1\>1_+\>1_+\}+\{\bar{\rm u}_b:1\>\bar1\>1\>1\>1_-\>1_-\}
= \{{\rm e}^-:1\>1\>\bar1\>\bar1\>1_+\>1_+\}+\{\bar\nu_e:\bar1 \>1\>1\>1\>\bar1_-\>\bar1_-\},\eqno (5.4)
$$
where the charge difference between the d and u quarks (determined by their values of C) 
corresponds to the charge difference between electrons and anti-neutrinos. 
This charge is carried by the weak bosons $\hat\gamma^{\pm}$, with 
the $Cl_{3,3}$ B,C description, given in \S2 of [2], 
$$
	\{\hat\gamma^+ :0\>\bar 2\} = \>{1\over  2}\,(\hat\gamma^{\pi7}+ i\hat\gamma^{\pi8}),\>\>\>\>\>\>
	\{\hat\gamma^- :0\> 2\}  = \>\>{1\over 2}\,(\hat\gamma^{\pi7}- i\hat\gamma^{\pi8}).   \eqno (5.5) 
$$

\beginsection \S6.  Conclusions

Following the discussion of parity and time-reversal in \S2, \S3 
develops new algebraic definitions, based on the CU formalism developed in [1,2].
This shows that coordinate parity and time direction correspond, respectively, to the  
quantum numbers C and B. Furthermore, C determines the intrinsic parity of fermions. 
\S4 relates the Standard Model interpretation CKM matrix to the C dependence of
the FG quantum numbers. This leads, in \S5, to modification of the 
Conservation Law, that states 
{\it all seven} quantum numbers, provided by the commuting elements of the $Cl_{7,7}$ 
algebra, are conserved in fermion decay processes and interactions. 
This law has direct applications in the prediction of processes that have 
yet to be observed. It should be of value in revised analyses of 
existing experimental results and the design of new experiments. 
	
\beginsection Appendix	
	
	$$\vbox
	{\settabs 10 \columns
		\+&{CDEF$_\pm$G$_\pm$ signatures for all four generations of fermions, with B=1}\cr 
		\+&|||||||||||||||||||||||||||||||||||||||||\cr
		\+&quark&C&D&E&F$_\pm$&G$_\pm$& $\rm Q_B$&$\rm Q_C$&Q\cr
		\+&|||||||||||||||||||||||||||||||||||||||||\cr
		\+&${\rm u}_b$    &$-1$   &$\>\>1$&$\>\>1$&$\>\>1_-$ &$\>\>1_-$  &1/6&1/2&$\>\>2/3$ \cr
		\+&${\rm u}_r$    &$-1$   &$\>\>1$&$-1$   &$\>\>1_-$ &$\>\>1_-$  &1/6&1/2&$\>\>2/3$ \cr
		\+&${\rm u}_g$    &$-1$   &$-1$   &$\>\>1$&$\>\>1_-$ &$\>\>1_-$  &1/6&1/2&$\>\>2/3$ \cr 
		\+ &&&\cr
		\+&${\rm d}_b$    &$\>\>1$&$\>\>1$&$\>\>1$&$\>\>1_+$ &$\>\>1_+$&1/6&$-1/2$&$-1/3$  \cr
		\+& ${\rm d}_r$   &$\>\>1$&$\>\>1$&$-1$   &$\>\>1_+$ &$\>\>1_+$&1/6&$-1/2$&$-1/3$  \cr
		\+&${\rm d}_g$    &$\>\>1$&$-1$   &$\>\>1$&$\>\>1_+$ &$\>\>1_+$&1/6&$-1/2$&$-1/3$  \cr 
		\+&|||||||||||||||||||||||||||||||||||||||||\cr
		\+&${\rm c}_b $   &$-1$   &$\>\>1$&$\>\>1$&$\>\>1_-$    &$-1_-$  &1/6&$1/2$&$\>\>2/3$\cr
		\+&${\rm c}_r $   &$-1$   &$\>\>1$&$-1$   &$\>\>1_-$    &$-1_-$  &1/6&$1/2$&$\>\>2/3$\cr
		\+& ${\rm c}_g $  &$-1$   &$-1$   &$\>\>1$&$\>\>1_-$    &$-1_-$  &1/6&$1/2$&$\>\>2/3$\cr 
		\+ &&&\cr
		\+& ${\rm s}_b $  &$\>\>1$&$\>\>1$&$\>\>1$& $\>\>1_+$   &$-1_+$  &1/6&$-1/2$& $-1/3$  \cr
		\+&${\rm s}_r $   &$\>\>1$&$\>\>1$&$-1$   &$\>\>1_+$    &$-1_+$  &1/6&$-1/2$& $-1/3$ \cr
		\+& ${\rm s}_g $  &$\>\>1$&$-1$   &$\>\>1$&$\>\>1_+$    &$-1_+$  &1/6&$-1/2$& $-1/3$  \cr
		\+&|||||||||||||||||||||||||||||||||||||||||\cr		
		\+ &${\rm t}_b $  &$-1$   &$\>\>1$&$\>\>1$&$-1_-$  &$\>\>1_-$   &1/6&$1/2$&$\>\>2/3$ \cr
		\+ &${\rm t}_r $  &$-1$   &$\>\>1$&$-1$   &$-1_-$  &$\>\>1_-$   &1/6&$1/2$&$\>\>2/3$ \cr
		\+ &${\rm t}_g $  &$-1$   &$-1$   &$\>\>1$&$-1_-$  &$\>\>1_-$   &1/6&$1/2$&$\>\>2/3$\cr 
		\+ &&&\cr
		\+ &${\rm b}_b $  &$\>\>1$&$\>\>1$&$\>\>1$&$-1_+$  &$\>\>1_+$   &1/6&$-1/2$&$-1/3$    \cr	
		\+ &${\rm b}_r $  &$\>\>1$&$\>\>1$&$-1$   &$-1_+$  &$\>\>1_+$   &1/6&$-1/2$&$-1/3$   \cr	
		\+ &${\rm b}_g $  &$\>\>1$&$-1$   &$\>\>1$&$-1_+$  &$\>\>1_+$   &1/6&$-1/2$&$-1/3$  \cr
		\+&|||||||||||||||||||||||||||||||||||||||||\cr	
		\+ &$\nu_e $       &$-1$&$-1$   &$-1$   &$\>\>1_-$ &$\>\>1_-$ &$-1/2$&1/2&$\>\>0$\cr	
		\+ &$\nu_\mu $     &$-1$&$-1$   &$-1$   &$\>\>1_-$ &$-1_-$    &$-1/2$&1/2&$\>\>0$\cr	
		\+ &$\nu_\tau $    &$-1$&$-1$   &$-1$   &$-1_-$    &$\>\>1_-$ &$-1/2$&1/2&$\>\>0$\cr
		\+ &&&\cr
		\+ &e$^-$       &$\>\>1$&$-1$   &$-1$   &$\>\>1_+$ &$\>\>1_+$& $-1/2$&$-1/2$&$-1$    \cr
		\+ &$\mu^-$     &$\>\>1$&$-1$   &$-1$   &$\>\>1_+$ &$-1_+   $&$-1/2$&$-1/2$&$-1$    \cr
		\+ &$\tau^-$    &$\>\>1$&$-1$   &$-1$   &$-1_+$    &$\>\>1_+$&$-1/2$&$-1/2$&$-1$    \cr 
		\+&|||||||||||||||||||||||||||||||||||||||||\cr	
		\+&$q_b(-4/3)$      &$-1$   &$\>\>1$   &$\>\>1$  &$-1$&$-1$ &1/6&$-3/2$&$-4/3$\cr
		\+&$q_r(-4/3)$      &$-1$   &$\>\>1$   &$-1$     &$-1$&$-1$ &1/6&$-3/2$&$-4/3$\cr
		\+&$q_g(-4/3)$      &$-1$   &$-1$      &$\>\>1$  &$-1$&$-1$ &1/6&$-3/2$&$-4/3$\cr
		\+ &&&\cr
		\+&$q_b(5/3)$      &$\>\>1$&$\>\>1$    &$\>\>1$   &$-1$&$-1$&1/6&3/2&$\>\>5/3$\cr 
		\+&$q_r(5/3)$      &$\>\>1$&$\>\>1$    &$-1$      &$-1$&$-1$&1/6&3/2&$\>\>5/3$\cr 
		\+&$q_g(5/3)$      &$\>\>1$&$-1$       &$\>\>1$   &$-1$&$-1$&1/6&3/2&$\>\>5/3$\cr 
		\+&|||||||||||||||||||||||||||||||||||||||||\cr	  	
		\+ &$l(-2)$        &$-1$&$-1$   &$-1$   &$-1$    &$-1$    &$-1/2$&$-3/2$&$-2    $\cr
		\+ &$l(1)$      &$\>\>1$&$-1$   &$-1$   &$-1$    &$-1$   &$-1/2$&$\>\>3/2$&$\>\>1$  \cr	
		\+&|||||||||||||||||||||||||||||||||||||||||\cr}
	$$
	Quantum numbers describing fourth generation fermions included in this table 
	are interpreted elsewhere [5].
	
	Fermion electric charges have two contributions, viz.
	$\rm Q_B ={1\over 6}( D + E- BDE)$ and $\rm Q_C=-{1\over 2}$( F + G - BFG)BC,
	giving total charges $\rm Q \times $(the magnitude of the electronic charge) with
	$$
	\rm Q = Q_B + Q_C ={1\over 6}( D + E- BDE) - {1\over 2}( F + G - BFG)BC.        \eqno (A1)
	$$
	This formula holds for both definitions of the F,G quantum numbers.
	Corresponding anti-fermion signatures and charges are obtained by reversing 
	the signs of all quantum numbers A to G. Quantum number conservation for 
	decays and interactions of fermions in the first three generations
	automatically conserve charge.

\beginsection References

\frenchspacing

\item {[1]} Newman, Douglas (2021) Unified theory of elementary fermions and their interactions
based on Clifford algebras {arXiv:2108.08274v11}

\item{[2]} Newman, D.J. (2024) Quantum number conservation: a tool in the design and 
analysis of high energy experiments. J.Phys.G: Nucl. Part. Phys. 51 095002 

\item {[3]} Thomson, Mark (2013) Modern Particle Physics (Cambridge University Press)

\item {[4]} Newman, Douglas (2021) Problems with the Standard Model of particle physics {arXiv:2308.12295v2} 

\item {[5]} Newman, Douglas (2022) Fourth generation fermions providing new candidates for Dark Matter
and Dark Energy {arXiv:2201.13238}

\end